\documentstyle[12pt,epsfig]{article}                                
\begin{document}

\titlepage                  

\begin{large}
\begin{bf}
 Spectral study on some complex PT-symmetry quantum systems using 
the Mathews-Lakshmanan position dependent mass  in von Roos model and derivation of one parameter  kinetic energy 
[ S.Cruz,J.Negro and L.M.Nieto [Phys.Lett $\bf{369} $(2007)]

\end{bf}

 \vspace{0.1cm}

 Biswanath Rath

\end{large}

\vspace{0.1cm}

$\dagger$ Department of Physics,
 Maharaja Sriram Chandra Bhanj Deo University,
 Takatpur, Baripada -757003, Odisha, INDIA

$\dagger$:biswanathrath10@gmail.com
\vspace{0.1cm}

$\bf{Abstract:}$

We investigate the unbroken and broken spectral nature  in some position 
dependent mass related to Mathews-Lakshmanan mass $m(x)=\frac{m_{0}}{(1+(\alpha x)^{2}}$ in  complex PT-symmetry  potentials with odd term in
 large$ x$ limit using the von Roos model kinetic energy following the suitabitable selection of two parameter condition $2a+b=1$[Cruz,Negro and Nieto:Phys.Lett. $\bf{A 369}$(2007)400].In fact the kinetic energy under the condition $b=2a$ becomes  
$T = -\frac{ 1}{m^{a}} \partial_{x}\frac{1}{m^{2a}}\partial_{x}\frac{1}{ m^{a}} $ a single parameter model, which can be derived using  a newly proposed momentum 
operator called
$\begin{bf} "pseudo-momentum \end{bf}$(P), where $P=-i \frac{1}{m^{a}} \partial_{x} \frac{1}{m^{a}} $. In fact, the spectra become unbroken in many suitable 
selection of  PT-symmetry i.e $V(x)\sim ix^{2K+1}$ (K=0,1,2).

\vspace{0.1cm}

\begin{bf}
PACS no-
\end{bf}

 03.65.Ge,

\vspace{0.1cm}

\begin{bf}
Key words:
\end{bf}
von Roos,two parameter condition,pseudo-momentum,unbroken spectra,broken spectra,Mathews-Lakshmanan non singular potential,PT-symmetry

\vspace{0.1cm}

\begin{bf}
1.Introduction
\end{bf}

Since 1983, effective mass model in semiconductor theory($Al_{x} Ga_{1-x}As$) has gained sufficient
 interest among potential authors to explore new physics coming out of the 
von Roos model 
[1,2]on  position dependent mass(PDM) Hamiltonian. Considering self-adjoint nature , the Mamiltonian can be written as    
\begin{equation}
H = \frac{1}{4}[ m(x)^{a}pm(x)^{b}pm(x)^{c}+m(x)^{c}pm(x)^{b}pm(x)^{c}] +  V(x)  \end{equation}
It is easy to check that 
\begin{equation}
 H = H^{\dagger}
\end{equation}
The  classical analogy of the above Hamiltonian is written as 
\begin{equation}
h=\frac{p^{2}}{2m(x)} + \zeta x^{2} \rightarrow \zeta=1 or 1/2
\end{equation}
In above the values of a,b,c are constrained subject to the condition 
$a+b+c=-1$. Some time back Cruz,Negro and Nieto[3] suggested that three parameter model can be reduced to two papameter model provided $a=c$.The prposed model 
now reduces to 
\begin{equation}
H = -\frac{1}{2}[ m(x)^{a}\frac{d}{ x} m(x)^{2b} \frac{d}{d x}m(x)^{a} + U(x)  \end{equation}
with $2a+b=-1$. In this two parameter Hamiltonian ,one can also have  large
 
no of choices. For example (i) $a=\frac{-1}{2};b=0$;(ii)$ b=1;a=0$;(iii) 

$a=\frac{-1}{\beta}$;$b=\frac{2-\beta}{\beta}$ with suitable choice for 

$\beta$. Simple model selected  by Cruz ,Negro and Nieto[3] is $\beta=4$. The Hamiltonian used earlier is 

\begin{equation}
H_{CNN} = \frac{1}{2}[-\frac{ 1}{m(x)^{1/4}} \partial_{x}\frac{1}{m^{1/2}}\partial_{x}\frac{1}{ m(x)^{1/4}} + V(x)]
\end{equation}

In fact authors have systematically selected few cases of singular and non-singular cases of mass. In fact the non-singular case of mass[3]
\begin{equation}
m(x)=\frac{m_{0}=1}{(1+  \alpha^{2} x^{2})}
\end{equation}
is commonly known as Mathews-Lakshmanan(ML)[4,5] which has been extensively studied classically by many authors. It is quite obvious that as the total Hamiltonian is Hermitian(self-adjoint) in nature, hence  the real unbroken spectra is
 guaranteed[5].
In fact analytical expressions on energy levels have been obtained[3,5].
 It is worth  mentioning  that position-dependent quantum systems still 
remains an open challenge in complex systems[6]. This motivated me to study
 the ML mass in complex space. In fact in complex space one of the primary criteria is that the Hamiltonian must be PT-invariant in nature[7]
\begin{equation}
[(PT),  H]=0
\end{equation}
where  
\begin{equation}
P x P^{-1} = -x
\end{equation}
\begin{equation}
P p P^{-1} = -p
\end{equation}
\begin{equation}
P |x| P^{-1} = |x|
\end{equation}
Similarly , the time reversal operator T has the following behaviour
\begin{equation}
T i T^{-1}= -i
\end{equation}
\begin{equation}
T p T^{-1} = -p
\end{equation}
\begin{equation}
T |p| T^{-1} = |p|
\end{equation}
However the term like $ix^{2K+1}$ is PT-symmetric
\begin{equation}
[ix^{2K+1},PT] = i x^{2K+1}
\end{equation}
\begin{equation}
[\frac{i}{x^{2K+1}} ,PT] = \frac{i}{x^{2K+1}}
\end{equation}
Similarly, the term like $x^{2}$ satisfies the following 
\begin{equation}
[x^{2},P]=0;[x^{2},T]=0;[x^{2},PT]=0
\end{equation}

This helps one to construct suitable models in PT-symmetry and study its spectra. In fact it should be borne in mind that PT-symmetry is not the sole criteria 
for real spectra. It may also exhibit complex spectra partially or fully.
However its study using ML type of mass is interesting and will be explored 
as given below. 

\begin{equation}
P x P^{-1} = -x
\end{equation}
\begin{equation}
P x P^{-1} = -x
\end{equation}

\begin{bf}
2.Model Kinetic  Energy Derivation using "Pseudo momentum"
\end{bf}

\begin{bf}
Two pameters Kinetic energy 
\end{bf}

Here we define a new term as "pseudo momentum "(P) 
which is related to momentum (p) as 

\begin{equation}
P=\frac{1}{m^{a}} p \frac{1}{ m^{b/2}}
\end{equation}
Then it is obvious that[8] 

\begin{equation}
P\neq P^{\dagger}
\end{equation}
then the kinetic energy in this case becomes[9] 
\begin{equation}
T=\frac{1}{2}[P P^{\dagger} + P^{\dagger} P]
\end{equation}

\begin{bf}
One parameter Kinetic energy 
\end{bf}

Here we adopt self-adjoint propertry on $P$ and define as 

\begin{equation}
P= \frac{1}{m^{a}} p \frac{1}{m^{a}}
\end{equation}
Hence we have 
\begin{equation}
P= P^{\dagger}
\end{equation}
which is like the standard form of momentum ($p$) ,which is also self-adjoint in its behaviour[8]
\begin{equation}
p= p^{\dagger}
\end{equation}
 Dimension wise both are different from each other but bearing the same
 self-adjoint property .
The corresponding kinetic energy becomes 
\begin{equation}
T = [-\frac{ 1}{m(x)^{a}} \partial_{x}\frac{1}{m^{2a}}\partial_{x}\frac{1}{ m(x)^{a}}]
\end{equation}
As the mass appears in denominator of kinetic energy we have $4a=1$,hence $a=\frac{1}{4}$.
The corresponding kinetic energy becomes 
\begin{equation}
T = [-\frac{ 1}{m(x)^{1/4}} \partial_{x}\frac{1}{m^{1/2}}\partial_{x}\frac{1}{ m(x)^{1/4}}]
\end{equation}

It should be remembered that for the K.E of the form 
\begin{equation}
T1 = \frac{1}{2}[-\frac{ 1}{m(x)^{1/4}} \partial_{x}\frac{1}{m^{1/2}}\partial_{x}\frac{1}{ m(x)^{1/4}}]
\end{equation}
The corresponding pseudo momentum can be written as 
\begin{equation}
P= \frac{1}{(2m)^{a}} p \frac{1}{(2m)^{a}}
\end{equation}
Now using above form of K.E (T) ,we will address the following model complex 
potentials as given below.

\begin{bf}
3. Position dependent Hamiltonian in complex space.
\end{bf}

At the outset ,we would like to say that all our equations are confined to commutation relation
\begin{equation}
[x,p]=-i[x,\partial_{x}]=-i
\end{equation}
so that inerested reader will understand the analysis appropriately.
 Here we consider mainly two types of model Hamiltonian using the same form of 
PDM.

\begin{bf}
case-I
\end{bf}
 
Here the Hamiltonian is of the form 
\begin{equation}
H =  [-\frac{ 1}{m(x)^{1/4}} \partial_{x}\frac{1}{m^{1/2}}\partial_{x}\frac{1}{ m(x)^{1/4}} +  i x^{2K+1}
\end{equation}
where$ K=0,1,2,...$

\begin{bf}
(i) K=0 : Broken
\end{bf}
The corresponding Hamiltonian becomes 
\begin{equation}
H^{(1)} =  [-\frac{ 1}{m(x)^{1/4}} \partial_{x}\frac{1}{m^{1/2}}\partial_{x}\frac{1}{ m(x)^{1/4}} +  i x 
\end{equation}

\begin{bf}
(ii) K=1: Unbroken 
\end{bf}

The corresponding Hamiltonian becomes 
\begin{equation}
H^{(2)} =  [-\frac{ 1}{m(x)^{1/4}} \partial_{x}\frac{1}{m^{1/2}}\partial_{x}\frac{1}{ m(x)^{1/4}} +  i x^{3} 
\end{equation}

\begin{bf}
(iii) K=2 : Unbroken 
\end{bf}

The corresponding Hamiltonian becomes 
\begin{equation}
H^{(3)} =  [-\frac{ 1}{m(x)^{1/4}} \partial_{x}\frac{1}{m^{1/2}}\partial_{x}\frac{1}{ m(x)^{1/4}} +  i x^{5} 
\end{equation}

\begin{bf}
(iv) K=3 :Unbroken
\end{bf}
The corresponding Hamiltonian becomes 
\begin{equation}
H^{(4)} =  [-\frac{ 1}{m(x)^{1/4}} \partial_{x}\frac{1}{m^{1/2}}\partial_{x}\frac{1}{ m(x)^{1/4}} +  i x^{7} 
\end{equation}

\begin{bf}
case-II
\end{bf}

\begin{equation}
H =  [-\frac{ 1}{m(x)^{1/4}} \partial_{x}\frac{1}{m^{1/2}}\partial_{x}\frac{1}{ m(x)^{1/4}} +  i m(x) x^{2K+1}
\end{equation}

\begin{bf}
(i) K=0 : Broken 
\end{bf}

The corresponding Hamiltonian becomes 
\begin{equation}
H^{(5)} =  [-\frac{ 1}{m(x)^{1/4}} \partial_{x}\frac{1}{m^{1/2}}\partial_{x}\frac{1}{ m(x)^{1/4}} +  im(x) x 
\end{equation}

\begin{bf}
(ii) K=1 : Broken
\end{bf}

The corresponding Hamiltonian becomes 
\begin{equation}
H^{(6)} =  [-\frac{ 1}{m(x)^{1/4}} \partial_{x}\frac{1}{m^{1/2}}\partial_{x}\frac{1}{ m(x)^{1/4}} +  im(x) x^{3} 
\end{equation}

\begin{bf}
(iii) K=2 : Unbroken
\end{bf}

The corresponding Hamiltonian becomes 
\begin{equation}
H^{(7)} =  [-\frac{ 1}{m(x)^{1/4}} \partial_{x}\frac{1}{m^{1/2}}\partial_{x}\frac{1}{ m(x)^{1/4}} +  i m(x)x^{5} 
\end{equation}

\begin{bf}
(iv) K=3 : Unbroken
\end{bf}
The corresponding Hamiltonian becomes 
\begin{equation}
H^{(8)} =  [-\frac{ 1}{m(x)^{1/4}} \partial_{x}\frac{1}{m^{1/2}}\partial_{x}\frac{1}{ m(x)^{1/4}} +  i m(x) x^{7} 
\end{equation}

\begin{bf}
4.Energy calculation method
\end{bf}

 Here we consider an exactly solvable model ,whose wave functions can be calculated exactly.The model is simple Harmonic Oscillator ,whose Hamiltonian is 

\begin{equation}
H_{0} = p^{2} + x^{2}
\end{equation}

whose energy eigenvalue satisfy the relation[10]
\begin{equation}
H_{0}|m>= (2m+1) |m>
\end{equation}
where 
\begin{equation}
|m>= N e^{-\frac{x^{2}}{2}} H_{m}(x) 
\end{equation}
In above N stands for normalization constant and $H_{m}(x)$ stands for Hermite polynomial.[8]
Here we use matrix diagonalisation method[5] to solve the energy eigenvalue relation
\begin{equation}
H_{p}|\Phi>=E|\Phi>
\end{equation}
where $|\Phi>$ is 
\begin{equation}
|\Phi>= \sum_{m} A_{m} |m>
\end{equation}

 As we have introduced a new term i.e "pseudo-momentum " we feel to calculate the values of this operator for the ground state i.e
\begin{equation}
<\Phi_{0}|P|\Phi_{0}>
\end{equation}
for all the unbroken energy states with a comparison with traditional momentum operator
\begin{equation}
<\Phi_{0}|P|\Phi_{0}>
\end{equation}

\begin{bf}
3. Result and Discussion
\end{bf}

Numerical values of energy levels are reflected graphically in figs.1-8. In fact in table-1, we cite the first four energy levels corresponding to unbroken spectra.In table-2 , the values of " pseudo momentum " operator have been reflected
for the ground state only. Lastly we would like to say that the von Roos[,2] K.E can be derived using the "pseudo-momentum concept". In the case of ML 
oscillator, unbroken energy levels depend on choice of K.

\pagebreak

\begin{table}

 Table-1: First four eigenvalues of position dependent mass Mathews-Lakshmanan  mass variation

\vspace{1.0cm}

\begin{tabular}{ c c  c  } \hline
quantum no   & Hamiltonian   & Energy levels  \\ \hline
0 & $H^{(1)}$ & 1.350 1  \\
1 & &4.815 1 \\
2& & 9.415 3 \\
3& & 14.917 6 \\  \hline
0 & $H^{(2)}$ & 1.467 2  \\
1 & &5.605 9 \\
2& & 11.8689 \\
3& & 19.835 8 \\  \hline
0 & $H^{(3)}$ &1.592 2  \\
1 & &6.224 5 \\
2& &13.558 0 \\
3& &23.247 9  \\  \hline
0 & $H^{(7)}$ & 1.145 5  \\
1 & &4.282 5 \\
2& & 9.415 3 \\
3& & 14.917 6 \\  \hline
0 & $H^{(8)}$ & 1.321 0  \\
1 & &5.120 9 \\
2& &11.022 8 \\
3& &18.661 6 \\  \hline
\end{tabular}

\end{table}

\begin{table}

 Table-2:Pseudo-momentum in von Roos model with Mathews-Lakshmanan position dependent mass

\vspace{1.0cm}

\begin{tabular}{ c c  c  } \hline
quantum no   & Hamiltonian   & $<\Psi_{0}|P|\Psi_{0}>$ \\ \hline
0 & $H^{(2)}$ & -0.558 0  \\
0 & $H^{(3)}$ & -0.467 5  \\
0 & $H^{(7)}$ & -0.579 2  \\
0 & $H^{(8)}$ & -0.479 7  \\  \hline
\end{tabular}

\end{table}

\pagebreak

\begin{figure}[h!]
\centering
\includegraphics{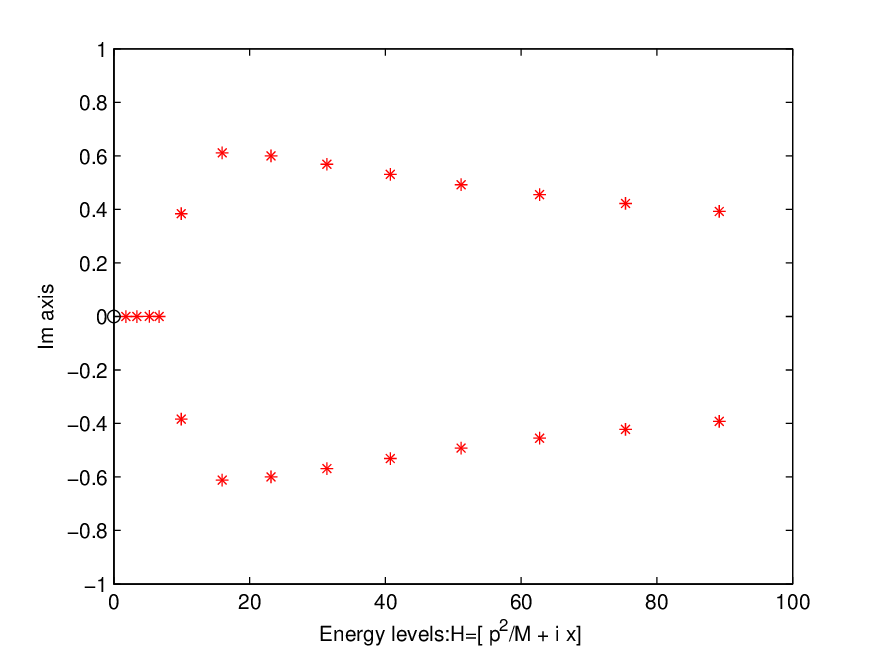}
\caption{ Energy levels of $H=\frac{p^{2}}{M}$ + ix$ :P M=\frac{1}{(1+x^{2})}$}
\end{figure}

\pagebreak

\begin{figure}[h!]
\centering
\includegraphics{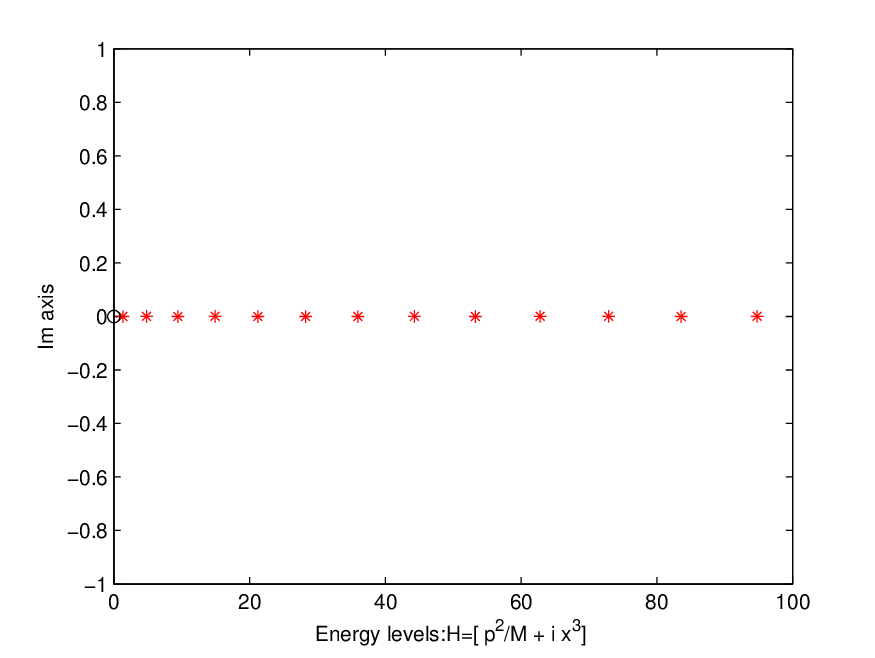}
\caption{ Energy levels of $H=\frac{p^{2}}{M} + ix^{3} :P M=\frac{1}{(1+x^{2})}$}
\end{figure}

\pagebreak

\begin{figure}[h!]
\centering
\includegraphics{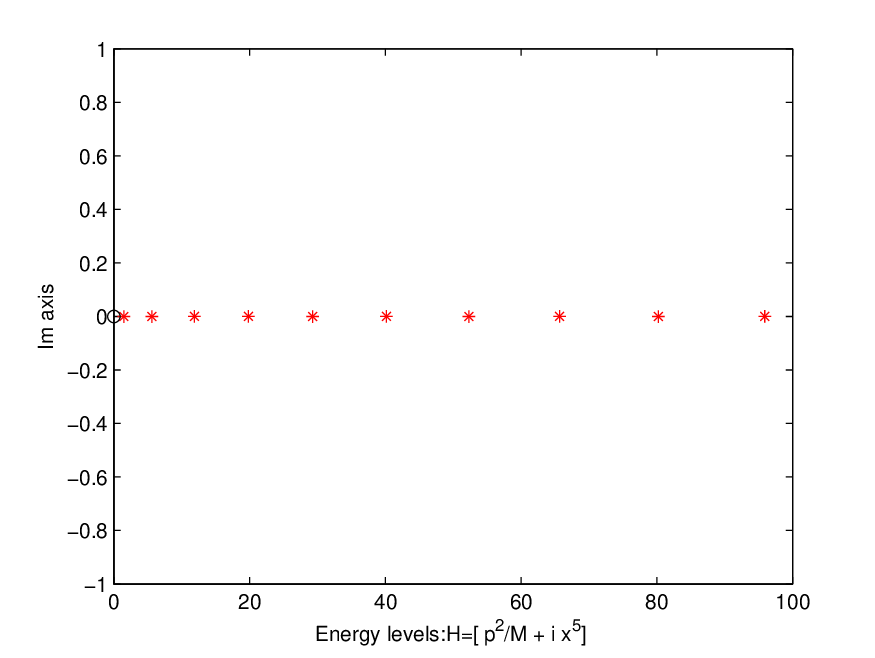}
\caption{ Energy levels of $H=\frac{p^{2}}{M} + ix^{5} :P M=\frac{1}{(1+x^{2})}$}
\end{figure}

\pagebreak

\begin{figure}[h!]
\centering
\includegraphics{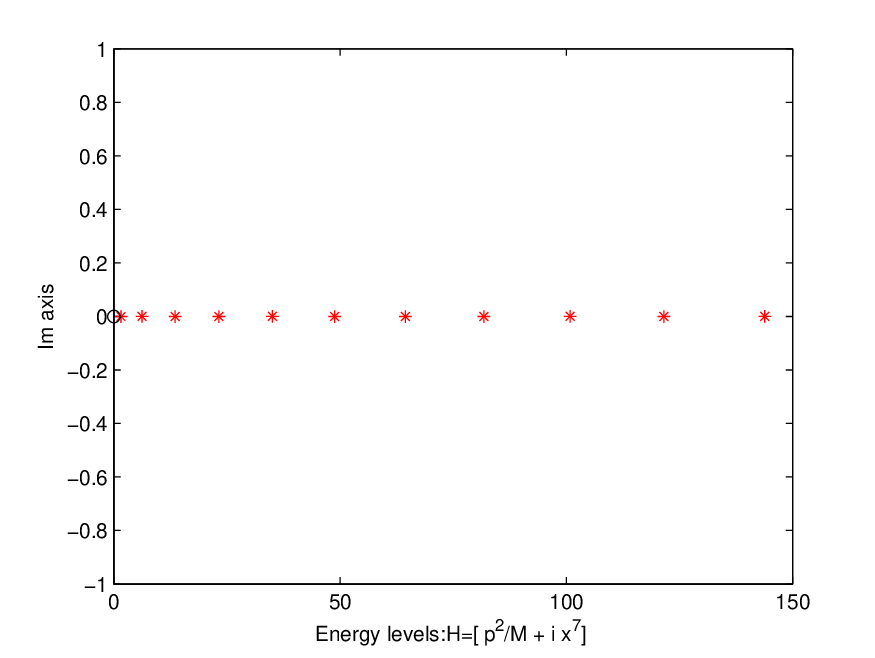}
\caption{ Energy levels of $H=\frac{p^{2}}{M} + ix^{7} :P M=\frac{1}{(1+x^{2})}$}
\end{figure}

\pagebreak

\begin{figure}[h!]
\centering
\includegraphics{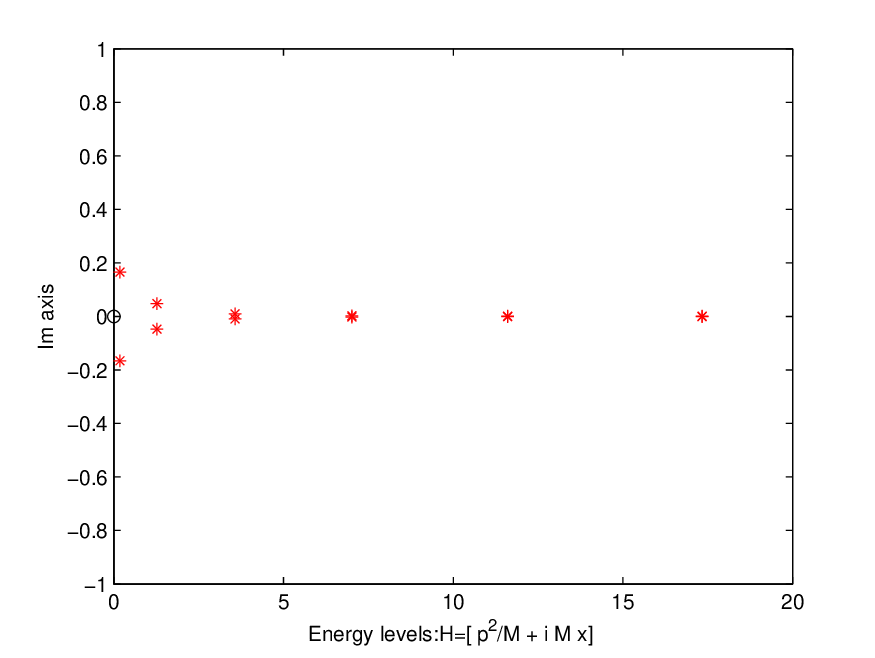}
\caption{ Energy levels of $H=\frac{p^{2}}{M} + M ix :P M=\frac{1}{(1+x^{2})}$}
\end{figure}

\pagebreak

\begin{figure}[h!]
\centering
\includegraphics{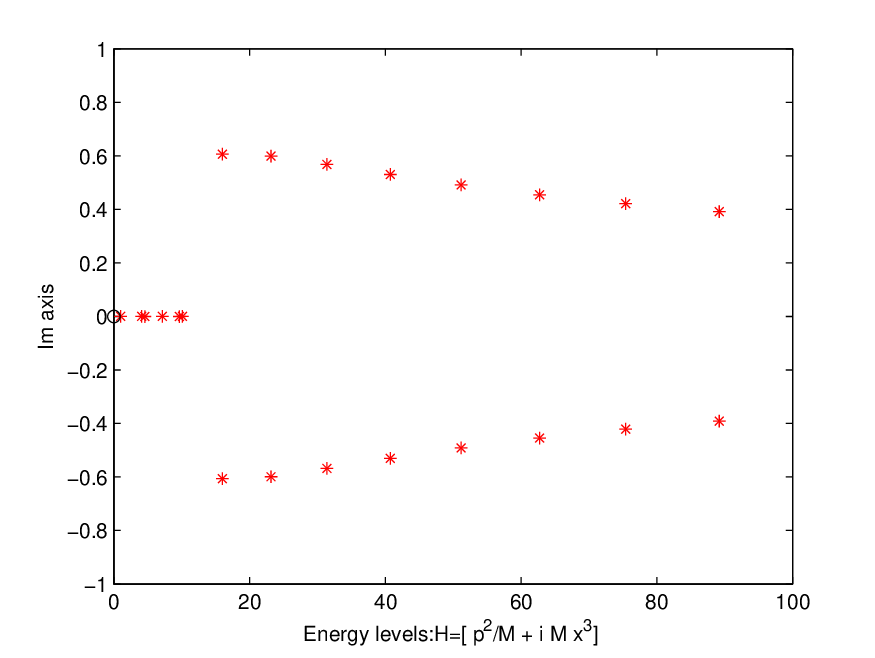}
\caption{ Energy levels of $H=\frac{p^{2}}{M} + iMx^{3} :P M=\frac{1}{(1+x^{2})}$}
\end{figure}
\pagebreak

\begin{figure}[h!]
\centering
\includegraphics{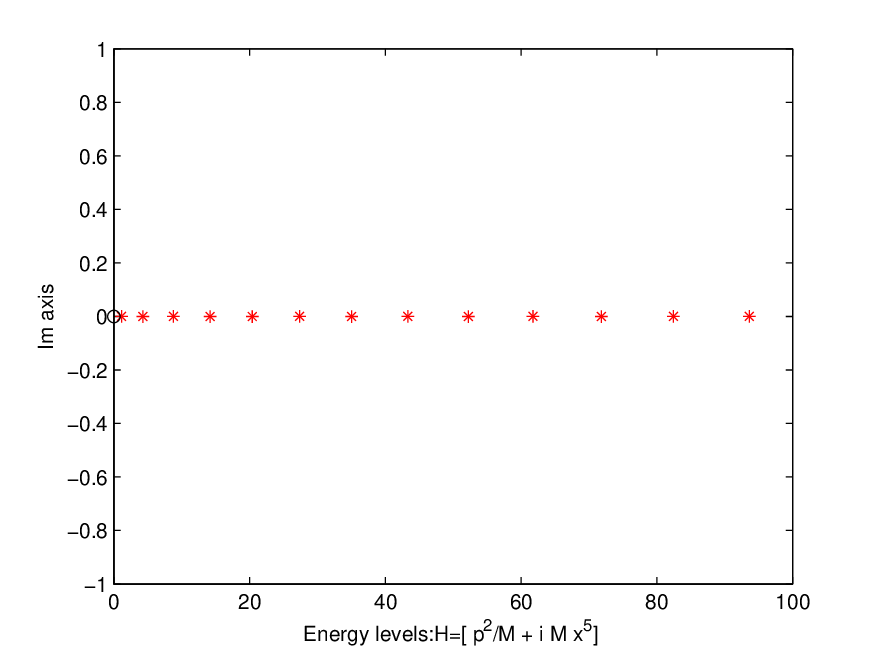}
\caption{ Energy levels of $H=\frac{p^{2}}{M} + iMx^{5} :P M=\frac{1}{(1+x^{2})}$}
\end{figure}
\pagebreak

\begin{figure}[h!]
\centering
\includegraphics{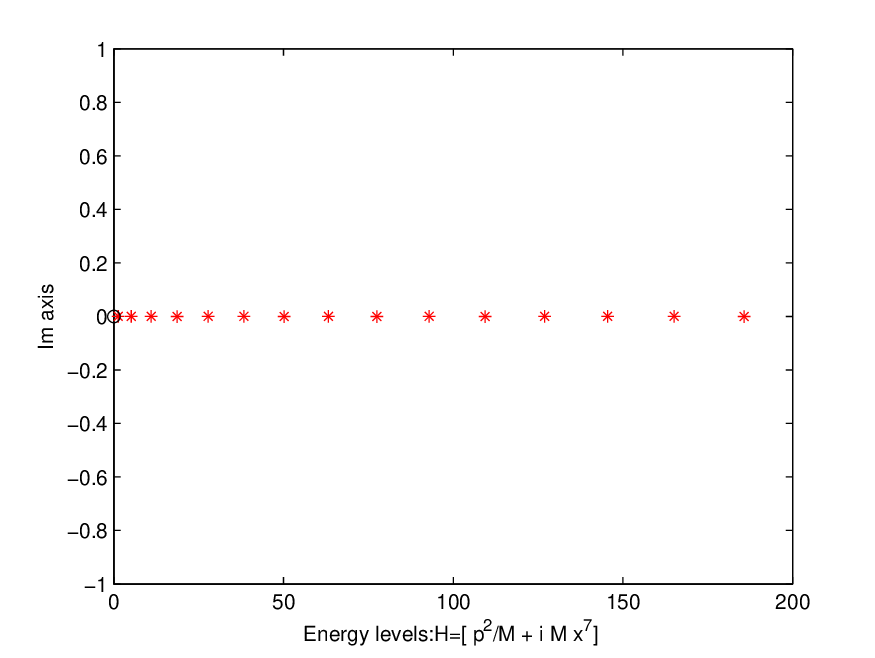}
\caption{ Energy levels of $H=\frac{p^{2}}{M} + iMx^{7} :P M=\frac{1}{(1+x^{2})}$}
\end{figure}


\begin{thebibliography}{99}

\bibitem{von}O.von Roos.Phys.Rev.$\bf{ B 27(12)}$.7547(1983).

\bibitem{von}O.von Roos and H.Mavromatis.Phys.Rev.$\bf{ B 31(4)}$.2294(1985).

\bibitem{Cruz}S.C.Cruz,J.Negro and L.M.Nieto, Phys.Lett $\bf{A 369}$,400(2007).

\bibitem{Mathews}P.M.Mathews and M.Lakshmanan,Quart.Appl.Math,$\bf{32}$(1974)215.
\bibitem{Buno}B.G.da Costa and E.P.Borges,J.Math.Phys $\bf{19}$(2018)042101.

\bibitem{Mustafa}O.Mustafa and S.habib.Mazharimousavi,Phys.Lett $\bf{A 373}$ (202009)325.


\bibitem{Bender}C.M.Bender,D.C.Brody and H.F.Jones,$\bf{89}$(2002)270401;$\bf{92}$(2004) 119902(E);C.M.Bender and H.F.Jones,Phys.Lett $\bf{A 328}$ (2004) 102.

\bibitem{yariv}A.Yariv"An Introduction to Theory and Applications of quantum mechanics" John Wiley and Sons,New York,(1982);D.J.Graffiths "Introduction to Quantum Mechanics" Second Edition,Pearson India,(2005);J.L.Powel and B.Crasemann,"Quantum Mechanics" Narosa,India(1988);J.J.Sakurai and J.J.Napolitano,Modern Quantum Mechanics",Pearson,India(2014);B.H.Bransden and C.J.Joachain,"Quantum Mechanics",Pearson India(2000);V.Rosansky,Introductory Quantum Mechanics",Asia Publishing House ,India(1962).

\bibitem{Rath}B.Rath,Phys.Scr $\bf{78(6)}$(2008)065012;$\bf{82(6)}$(2010) 069801(Addendum.
\bibitem{Rath}B.Rath and H.Mavromatis.Ind.J.Phys.$\bf{ B73(4)}$,641(1999);
B.Rath.Eur.Phys.J.Plus $\bf{136}$.493(2021);B.Rath and P.Mahapatra.Results.in.Phys $\bf{25}$(2021)104197.

\end{thebibliography}
\end{document}